\documentclass[epj,twocolumn]{webofc} 
\usepackage[varg]{txfonts} 
\usepackage{graphics}
\begin{document}
\title{LHC data and cosmic ray coplanarity at superhigh energies}

\subtitle{}

\author{Rauf Mukhamedshin\inst{1}\thanks{\email{rauf_m@mail.ru}}
}


\institute{Institute for Nuclear Research, Russian Academy of Sciences, Moscow, Russia
}

\abstract{A new phenomenological model FANSY 2.0 is designed, which makes it possible to simulate hadron interactions via
traditional and coplanar generation of most energetic particles as well as to reproduce a lot of LHC (ALICE,
ATLAS, CMS, TOTEM, LHCf) data.
  Features of the model are compared with LHC data. Problems of coplanarity are considered and a testing experiment is proposed.
} 

\maketitle

\section{Introduction}
\label{intro}
A number of models of hadron interactions with nuclei is concurrently applied in cosmic-ray experiments as none of them can explain the entire experimental data set of EAS features. Besides, a number of phenomena observed in mountain-based
and stratospheric X-ray--emulsion chamber (XREC)  experiments is not explained yet.
 One of these interesting phenomena is the so-called coplanarity of most energetic cores of $\gamma$-ray--hadron families, i.e., groups of high-energy ($E\gtrsim n\cdot 1$ TeV) particles in relatively young EAS cores
initiated by protons and nuclei of the primary cosmic radiation (PCR) (see Sect. 3).

The phenomenological model FANSY 1.0 has been designed a few years ago, which helped to understand general features of coplanar events \cite{mukhEPJC2009}.
FANSY 1.0 gives reasonable results as compared with experimental data on intensity of $\gamma$-ray families, energy dependence of the muon and hadron spectra  and so on \cite{mukhEPJC2009}.

However, FANSY 1.0 cannot properly reproduce LHCf data on high-$X_F \gamma$-rays (\cite{LHCfPLB2011}, e.g.) and neutrons, which are really important for cosmic ray experiments, as well as  the transversal size of $\gamma$-ray families, e.g. \cite{KemPMukhTam2013}.

To improve this situation, a new phenomenological model FANSY 2.0 is designed, which  simulates traditional and coplanar particle generation and reproduce experimental data with a higher accuracy as compared with FANSY 1.0. FANSY 2.0 reproduces a number of ALICE, ATLAS, CMS, TOTEM data
on charged, strange and charmed stable and resonance particle generation, LHCf data on high-X$_F$ gamma-rays and neutrons.

This paper is devoted to $pp$ interactions simulated at superhigh energies $(\sqrt{s}=900$ GeV $- 13$ TeV).
 Compatibility of FANSY 2.0 and LHC data will be discussed in more detail in another work \cite{mukhcopl}

This paper is organized as follows. Sect. 2 presents comparison of LHC data and simulated results. Coplanarity problem is briefly considered in Sect. 3. Final comments are given in Conclusion.


\section{High energy interactions}

\subsection{Cross sections}
\label{general}

\begin{figure}[ht]
\centering
   \includegraphics[width=0.48\textwidth]{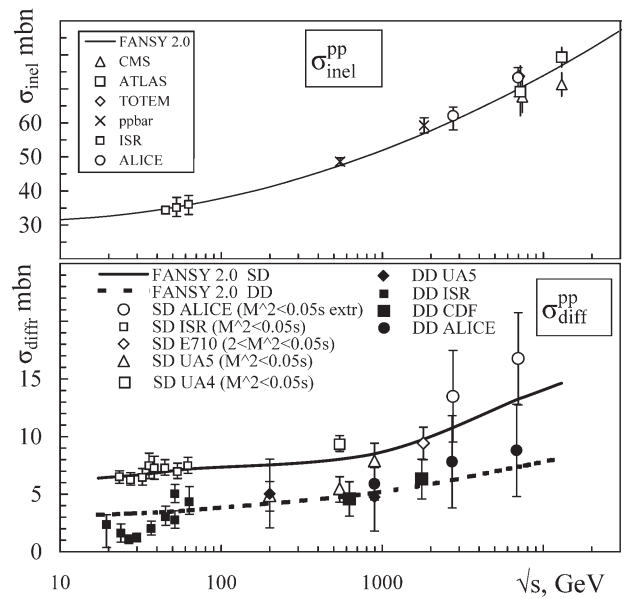}
  \caption{Experimental (points) and FANSY 2.0's (curve) energy dependencies of inelastic $pp$ cross section (top); single and double diffraction diffraction sections (bottom).}
  \label{pp_crosssects}
\end{figure}

Fig. \ref{pp_crosssects} shows FANSY 2.0's results (curve) and data of different experiments on energy dependence of the inelastic $pp$ cross section $\sigma_{inel}^{pp}$ (top);  single and double diffraction  sections, $\sigma_{SD}^{pp}$ and $\sigma_{DD}^{pp}$ (bottom). Inelastic $pp$ cross section data include results of the ATLAS \cite{arXiv1606.02625} and CMS \cite{arXiv1607.02033} Collaborations at $\sqrt{s} =  13$ TeV.

\subsection{Low-(pseudo)rapidity data}

Experimental LHC data are mainly related to low-(pseudo)rapidity
particles. This kinematic range is not very important for consideration of cosmic-ray-initiated EAS's in the atmosphere. However, models, which cannot describe these data, will inevitably cause mistrust. Experimental and simulated data are shown in Sect. 2  with filled and empty symbols, respectively.

\subsubsection{Charged particles}

\begin{figure}[ht]
\centering
   \includegraphics[width=0.48\textwidth]{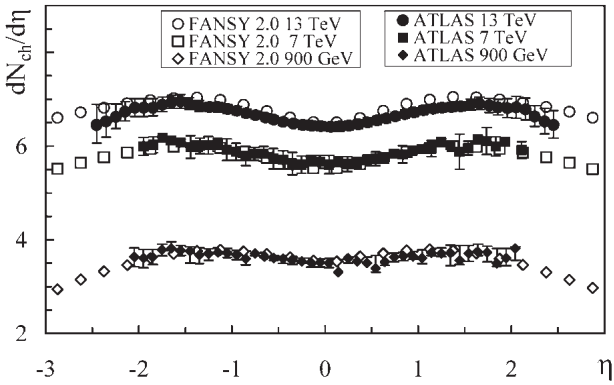}
  \caption{FANSY 2.0 and ATLAS $dN_{ch}/d\eta$ distributions of charged particles at $-2.1 < \eta < 2.1$, $n_{ch} \geq 2$, and $p_t > 0.1$ GeV/c in NSD $pp$ interactions at $\sqrt{s} = 13$ TeV (circles), 7 TeV (squares) and 900 GeV (diamonds).}
  \label{dnchdetaATLAS}
\end{figure}

Fig. \ref{dnchdetaATLAS} shows ATLAS $dN_{ch}/d\eta$ distribution of charged particles at $-2.1 < \eta < 2.1$, $n_{ch} \geq 2$, and $p_t > 0.1$ GeV/c in NSD $pp$ interactions at $\sqrt{s} = 13$ TeV (circles) \cite{ATLASdNchdEta13TeV}, 7 TeV (squares) and 900 GeV (diamonds) \cite{ATLASdNchdEta0.9_7TeV}.

\begin{figure}[ht]
\centering
  \includegraphics[width=0.48\textwidth]{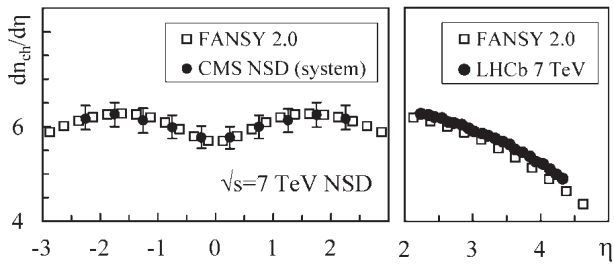}
  \caption{FANSY 2.0 and experimental $dN_{ch}/d\eta$ distributions of charged particles in $pp$ NSD interactions at $\sqrt{s} = 7$ TeV of the CMS experiment at $-2.5 < \eta < 2.5$ (left) and LHCb experiment at $2.0 < \eta < 4.5$ (right).}
  \label{dnchdetaCMSLHCb7TeV}
\end{figure}

Fig. \ref{dnchdetaCMSLHCb7TeV} shows experimental $dN_{ch}/d\eta$ distributions of charged particles in $pp$ NSD interactions at $\sqrt{s} = 7$ TeV of the CMS experiment \cite{dnchdetaCMS1005.3299} at $-2.5 < \eta < 2.5$ (left) and the LHCb experiment at $2.0 < \eta < 4.5$ (right) \cite{LHCbdNchdEta7TeV}.

\subsubsection{Strange particles}

\begin{figure}[ht]
\centering
   \includegraphics[width=0.48\textwidth]{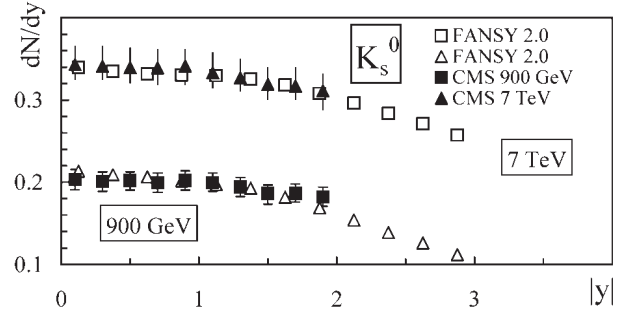}
  \caption{CMS $dN/d|y|$ distributions of K$^0_s$ mesons in $pp$ NSD interactions at  $\sqrt{s} = 900$ GeV (lower squares) and $\sqrt{s} = 7$ TeV  (upper triangles).}
  \label{pp_Ks_LHC}
\end{figure}

Fig. \ref{pp_Ks_LHC} shows FANSY 2.0  and CMS $dN/d|y|$ distributions of K$^0_s$ mesons in $pp$ NSD interactions at  $\sqrt{s} = 900$ GeV (lower squares) and $\sqrt{s} = 7$ TeV  (upper triangles) \cite{JHEP05_2011_064}.

\begin{figure}[ht]
\centering
   \includegraphics[width=0.48\textwidth]{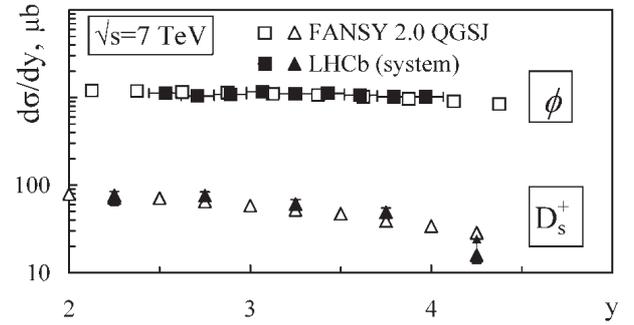}
  \caption{FANSY 2.0 and LHCb $d\sigma /dy$ distributions of $\phi$ mesons at $0.6 < p_t < 5.0$ GeV/c (upper squares); D$^\pm_s$ mesons at $1.0 < p_t < 8.0$ GeV/c (lower triangles) at $\sqrt{s} = 7$ TeV.}
  \label{ppphiDschdNdetaLHCb}
\end{figure}

Fig. \ref{ppphiDschdNdetaLHCb} shows FANSY 2.0  and LHCb experiment's $d\sigma /dy$ distributions of $\phi$ mesons with $0.6 < p_t < 5.0$ GeV/c at  $\sqrt{s} = 7$ TeV (upper squares) \cite{PhysLettB703}.

\subsubsection{Charmed particles}

Among different aims, FANSY 2.0 is designed to study "forward-physics"  aspects of
generation of charmed particles. So these particles are considered below in more details
as compared with light-quark hadrons.

Fig. \ref{ppphiDschdNdetaLHCb} shows LHCb $d\sigma/dy$ distribution of D$^\pm_s$ mesons with $1.0 < p_t < 8.0$ GeV/c  at $\sqrt{s} = 7$ TeV in $pp$ interactions (lower triangles).

\begin{figure}[ht]
\centering
   \includegraphics[width=0.48\textwidth]{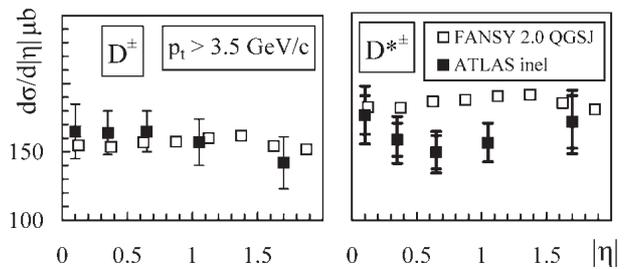}
  \caption{ATLAS and  FANSY 2.0's $d\sigma/d|\eta|$ distribution of D$^\pm$ mesons (left) and D*$^\pm$ mesons  (right)  with $p_t > 3.5$ GeV/c.}
  \label{ppDchdndetaATLAS}
\end{figure}

 Fig. \ref{ppDchdndetaATLAS} shows FANSY 2.0  and ATLAS $d\sigma/d|\eta|$  distributions of D$^\pm$ (left) and  D*$^\pm$ (right) mesons with $p_t > 3.5$ GeV/c in $pp$ interactions at $\sqrt{s} = 7$ TeV \cite{arXiv1512.02913v1}.

\begin{figure}[ht]
\centering
   \includegraphics[width=0.48\textwidth]{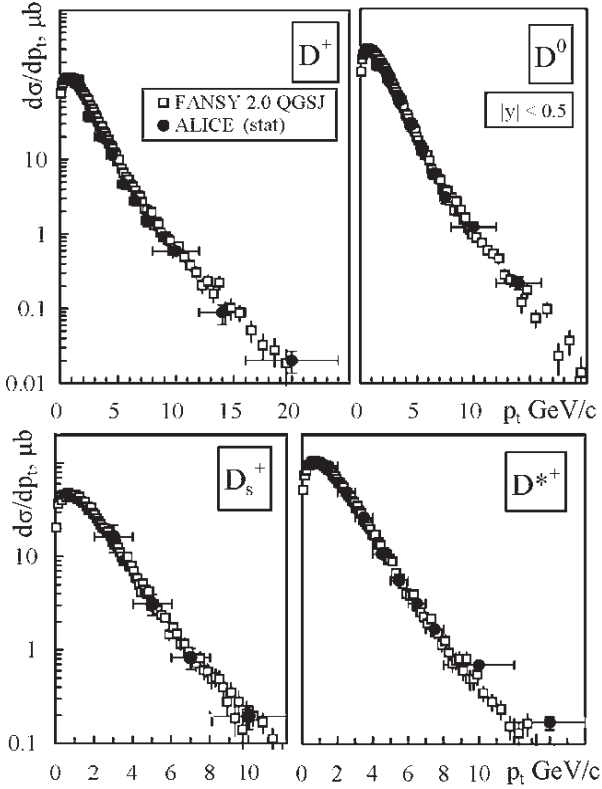}
  \caption{ALICE  $d\sigma /dp_t$ distributions of  D$^+$ (top,left), D$^0$ (top,right) mesons as well as strange
 D$_s^+$ (bottom,left), vector D$^{*+}$  (bottom,right) mesons
 at $|y|< 0.5$.}
  \label{ppDmesonsdsdpt7TALICE}
\end{figure}

Fig. \ref{ppDmesonsdsdpt7TALICE}  shows  ALICE $d\sigma/dp_t$  distributions of
 D$^+$ (top,left), D$^0$ (top,right) mesons as well as strange
 D$_s^+$ (bottom,left), vector D$^{*+}$  (bottom,right) mesons
 at $|y|< 0.5$ and $\sqrt{s} = 7$ TeV in $pp$ interactions \cite{JHEP01_2012_128}. Experimental system uncertainties are approximately equal to statistical ones.

\begin{figure}[ht]
\centering
   \includegraphics[width=0.48\textwidth]{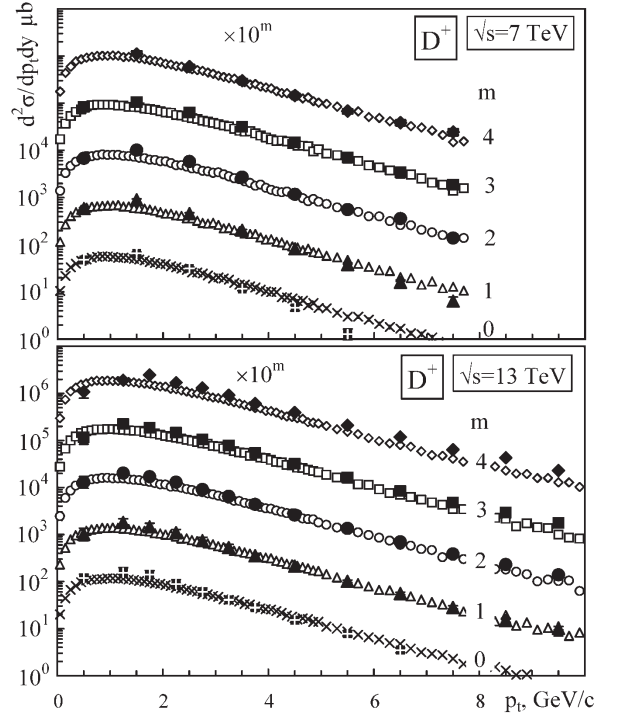}
\caption{LHCb $d^2\sigma /dy dp_t$  distributions  of D$^0$ mesons
in five rapidity regions ($2.0 - 2.5$, $2.5 - 3.0$, $3.0 - 3.5$, $3.5 - 4.0$, $4.0 - 4.5$) at $\sqrt{s} = 7$ TeV (top) and $\sqrt{s} = 13$ TeV
(bottom) with filled diamonds, squares, circles, triangles, and crosses, respectively.
Spectra are multiplied by $10^{m}$, where $m=4,3,2,1,0$ for the above-mentioned $y$ ranges,
respectively.}
  \label{dptpD+7_13LHCb}
\end{figure}

Fig. \ref{dptpD+7_13LHCb} shows LHCb data on $d^2\sigma /dp_tdy $ distributions of D$^+$ mesons in five rapidity regions, namely, at $2.0 - 2.5$, $2.5 - 3.0$, $3.0 - 3.5$, $3.5 - 4.0$, and $4.0 - 4.5$ at $\sqrt{s} = 7$ TeV (top) and $\sqrt{s} = 13$ TeV (bottom) with diamonds, squares, circles, triangles,
and crosses, respectively  \cite{LHCbcharm7TeV,LHCbcharm13TeV}. Spectra are multiplied by $10^{m}$, where $m=4,3,2,1,0$ for the
above-mentioned rapidity ranges, respectively.

\begin{figure}[ht]
\centering
   \includegraphics[width=0.48\textwidth]{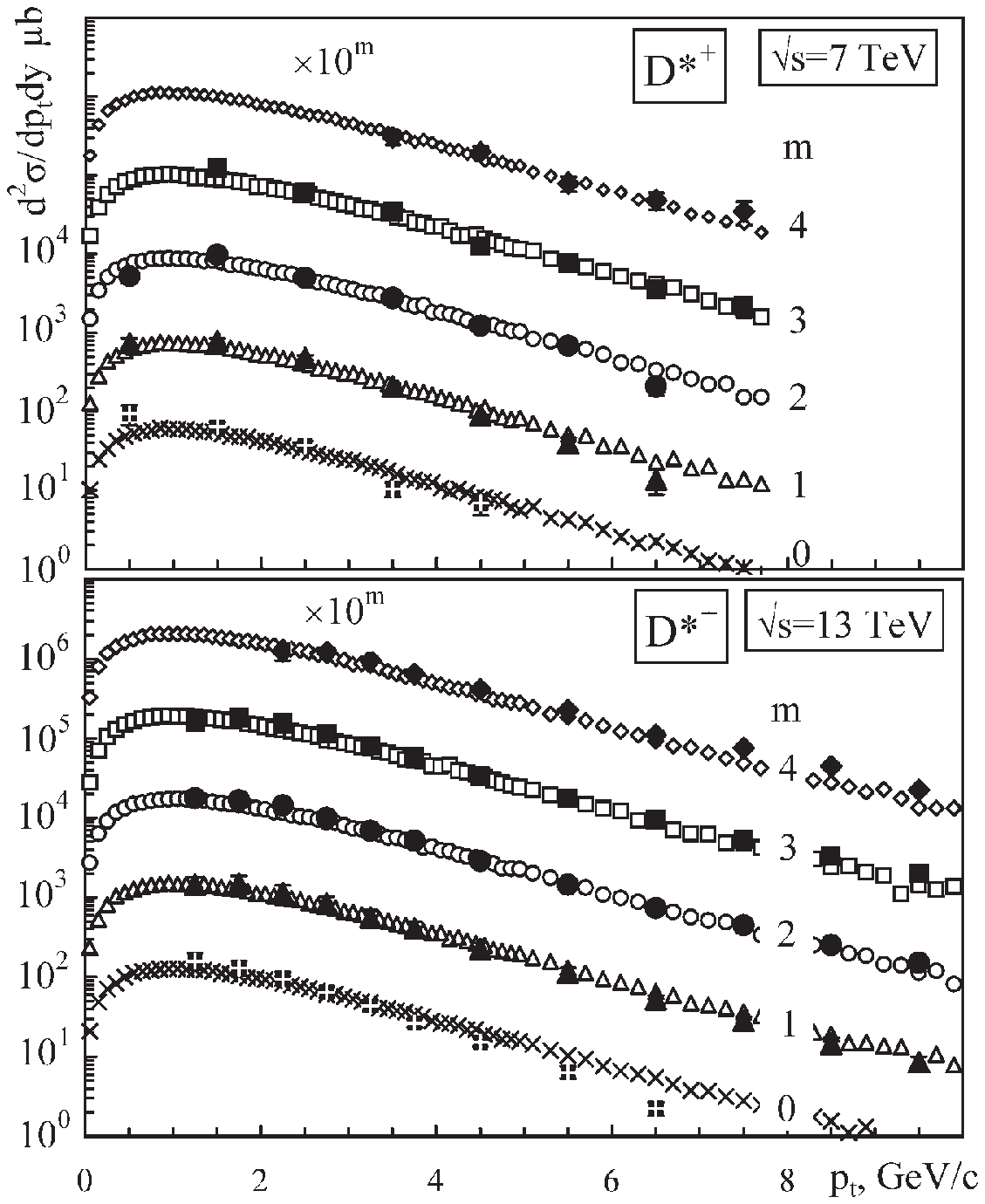}
\caption{LHCb $d^2\sigma /dp_tdy $ distributions of D$^{*+}$ mesons in five rapidity regions. Notations are the same as in Fig. \ref{dptpD+7_13LHCb}.}
  \label{dptpDstar+7_13LHCb}
\end{figure}

\begin{figure}[ht]
\centering
   \includegraphics[width=0.48\textwidth]{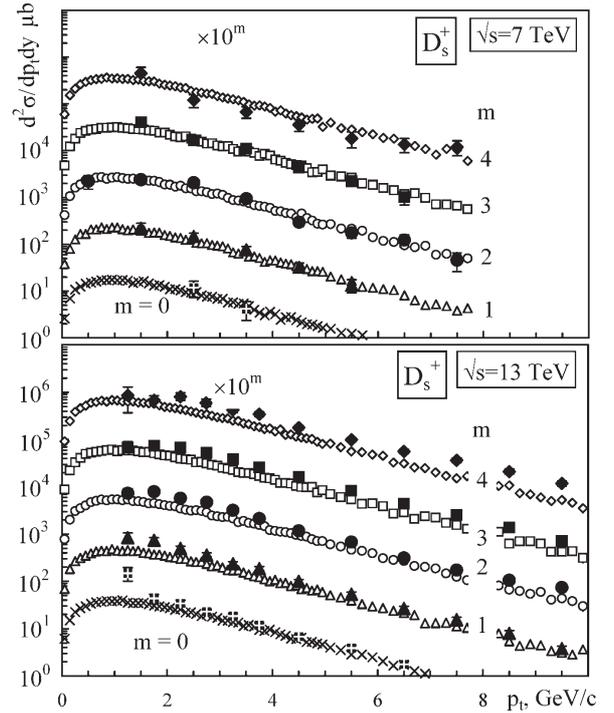}
\caption{LHCb $d^2\sigma /dp_tdy $ distributions of D$_s^{+}$  mesons in five rapidity regions. Notations are the same as in Fig. \ref{dptpD+7_13LHCb}.}
  \label{dptpDs+7_13LHCb}
\end{figure}

Figs. \ref{dptpDstar+7_13LHCb} and  \ref{dptpDs+7_13LHCb} show LHCb $d^2\sigma /dp_tdy $ distributions of D$^{*+}$ and
D$_s^{+}$ mesons, respectively \cite{LHCbcharm7TeV,LHCbcharm13TeV}. All the notations are the same as in Fig. \ref{dptpD+7_13LHCb}.

\begin{figure}[ht]
\centering
   \includegraphics[width=0.48\textwidth]{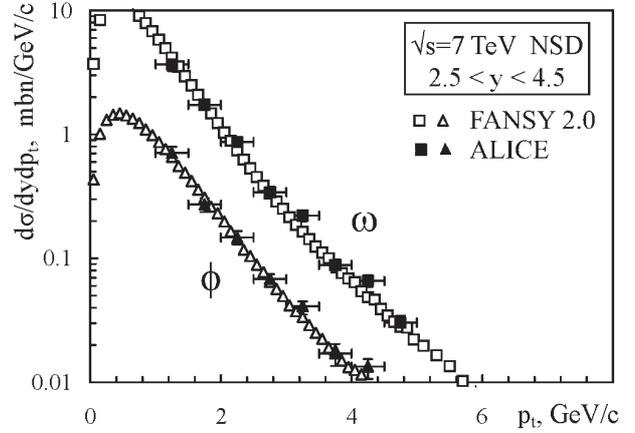}
  \caption{FANSY 2.0 and ALICE $d^2\sigma /dy dp_t$ distributions of $\omega^{0}$ (upper squares) and $\phi$  (lower triangles) mesons, respectively, at  $2.5 < y < 4.5$ at $\sqrt{s} = 7$ TeV .}
  \label{ppomegaphidsdpt7TALICE}
\end{figure}

 Fig. \ref{ppomegaphidsdpt7TALICE} shows data of the ALICE experiment on $d^2\sigma /dy dp_t$ of $\omega^{0}$ (squares) and $\phi$ mesons (triangles) at $2.5 < y < 4.5$ \cite{omegaphidsdptALICE}.
  Comparison of cross sections of generation of these mesons gives information on suppression of strange quark generation.

\begin{figure}[ht]
\centering
   \includegraphics[width=0.48\textwidth]{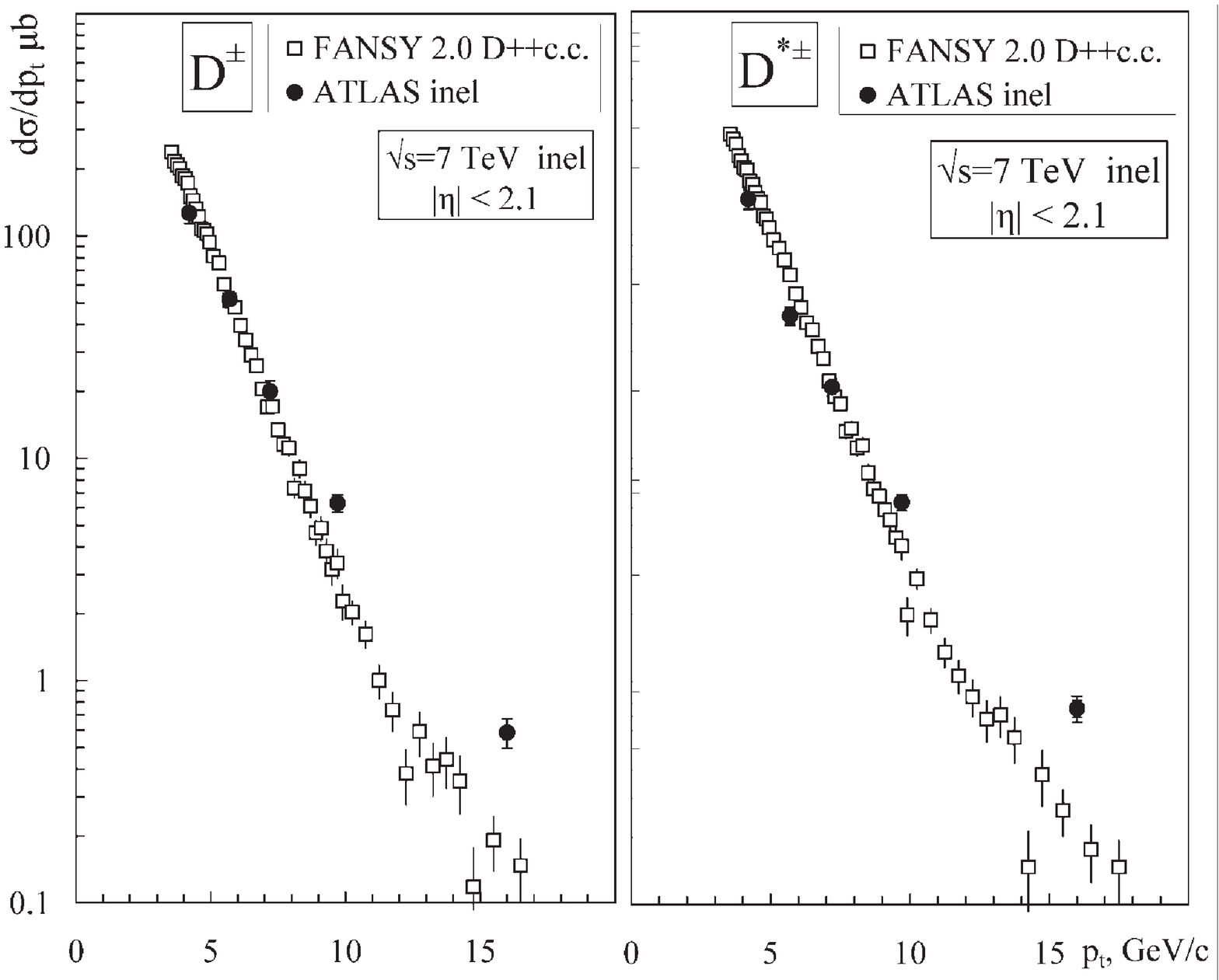}
\caption{ATLAS $d\sigma /dp_t$ spectra of D$^{\pm}$ (left) and D$^{*\pm}$ (right) mesons at
$|\eta|< 2.1$ and $\sqrt{s} = 7$ TeV.}
  \label{dptpD+Dstar+ATLAS}
\end{figure}

Fig. \ref{dptpD+Dstar+ATLAS} shows experimental $d\sigma /dp_t$ spectra of  D$^{\pm}$ (left) and D$^{*\pm}$ (right) mesons at $|\eta|< 2.1$ by the ATLAS experiment at $\sqrt{s}= 7$ TeV.

Table \ref{Dcrosssect} shows ALICE and FANSY 2.0 cross sections of  D$^0$, D$^+$ and D$^{*+}$  charmed-meson generation at $|\eta |<0.5$ in wide $p_t$ ranges.

Table \ref{DtoDratios} shows ALICE and FANSY 2.0 ratios of yields of charmed mesons, $\textrm{D}_s^+/$D$^+$ and $\textrm{D}_s^+/$D$^0$, as well as a ratio of prompt vector  mesons to prompt vector+pseudoscalar mesons, $P_V= \textrm{D}^*/(\textrm{D}^{*+}+\textrm{D}^+)$,  at $|\eta |<0.5$.

\begin{table}
\caption{\label{Dcrosssect} D$^0$, D$^+$ and D$^{*+}$  charmed-meson generation cross sections (mbarn).  ALICE data include statistical and systematical errors. FANSY 2.0 data include statistical errors.}
\begin{tabular}{lccc}
\hline\noalign{\smallskip}

   Particles                &  $p_t$,    &  ALICE                         & FANSY 2.0  \\
                            &  GeV/c     & $\sigma \pm$ stat. $\pm$ syst. & $\sigma \pm$ stat. \\
\hline\noalign{\smallskip}
 $\textrm{D}^0$             &   $1-16$   & $412 \pm 33\genfrac{}{}{0pt}{}{+55}{-140} $  & $457 \pm 6$ \\
 $\textrm{D}^+ $            &   $1-24$   & $198 \pm 24\genfrac{}{}{0pt}{}{+42}{-73}  $  & $201 \pm 4$ \\
 $\textrm{D}^{*+}$          &   $1-24$   & $203 \pm 23\genfrac{}{}{0pt}{}{+30}{-67}  $  & $180  \pm 2$ \\
\noalign{\smallskip}\hline
\noalign{\smallskip}
\end{tabular}
\end{table}

\begin{table}
\caption{\label{DtoDratios} $\textrm{D}_s^+/$D$^+$, $\textrm{D}_s^+/$D$^0$ ratios and ratio of vector mesons to prompt vector+pseudoscalar ones, $P_V$, at $|\eta |<0.5$. ALICE data include statistical and systematical errors. FANSY 2.0 results include statistical errors.}
\begin{tabular}{@{}*{4}{c}}
\hline\noalign{\smallskip}

   Particles                      &  $p_t$,        & ALICE           & FANSY 2.0 \\
                                  &   GeV/c        & & \\
\hline\noalign{\smallskip}
 $\textrm{D}_s^+ / \textrm{D}^+$  &   $2-12$       & $0.36 \pm 0.11\pm 0.12 $  & $0.35 \pm 0.01$ \\
 $\textrm{D}_s^+ / \textrm{D}^0$  &   $2-12$       & $0.20 \pm 0.05\pm 0.06 $  & $0.15 \pm 0.01$  \\
 $P_V= \textrm{D}^{*+}/$ &&&\\
 $(\textrm{D}^{*+}+\textrm{D}^+)$  &  $1-24$        & $0.59 \pm 0.06\pm 0.08 $   & $0.55 \pm 0.01$ \\
\noalign{\smallskip}\hline\noalign{\smallskip}
\end{tabular}
\end{table}

\subsubsection{High $\eta$ and $X_F$ data}

Experimental high $\eta$ and $X_F$ data are more interesting for cosmic ray experiments.
 Unfortunately, at the present time the specific design of colliders makes it possible to derive highest $\eta$ and $X_F$ data for low-$p_t$ neutral particles only.

\begin{figure*}[tb]
   \includegraphics[width=0.99\textwidth]{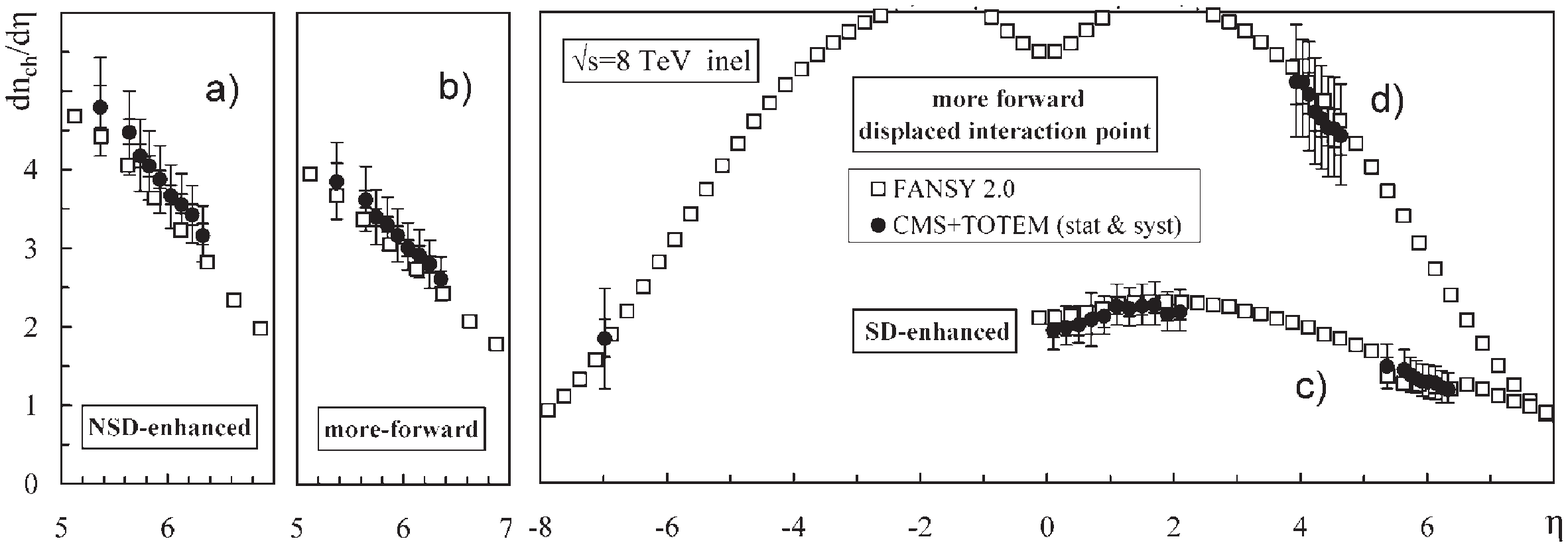}
  \caption{CMS+TOTEM and FANSY 2.0's $dn_{ch}/d\eta$ data for events selected at $\sqrt{s} = 8$ TeV under the following requirements.  "NSD-enhanced" data: $n_{ch} \geq 1$ in the ranges $-6.5 < \eta <-5.3$ \textbf{and} $5.3 < \eta < 6.5$ (a);  "more-forward" data:  $n_{ch} \geq 1$ in the ranges $-6.5 < \eta <-5.3$ \textbf{or}  $5.3 < \eta < 6.5$ (b);  "SD-enhanced" data (lower triangles): $n_{ch} \geq 1$ \textbf{only} in the ranges $-6.5 < \eta <-5.3$ \textbf{or only} $5.3 < \eta < 6.5$ (c);  "more-forward" data (upper squares) derived with displaced interaction points: $n_{ch} \geq 1$  in the ranges $-7.0 < \eta <-6.0$ \textbf{or} $3.7 < \eta < 4.8$ (d).}
  \label{CMSTOTEM}
\end{figure*}

Data of the CMS+TOTEM experiment \cite{CMSTOTEM} at $\sqrt{s} = 8$ TeV are of especial interest.
 Fig. \ref{CMSTOTEM} shows "NSD-enhanced" $dn_{ch}/d\eta$ distributions
 for events selected at $\sqrt{s} = 8$ TeV under the following requirements. Number of detected charged particles
$n_{ch} \geq 1 $ at $-6.5 < \eta <-5.3$ \textbf{and} $5.3 < \eta < 6.5$ (a);  "more-forward" data for events with $n_{ch} \geq 1$ at $-6.5 < \eta <-5.3$ \textbf{or}  $5.3 < \eta < 6.5$ (b); "SD-enhanced" data derived for events with $n_{ch} \geq 1 $ $only$ at $-6.5 < \eta <-5.3$  \textbf{or only} $5.3 < \eta < 6.5$  (lower triangles) (c);  "more-forward" data obtained with displaced interaction points for events with  $n_{ch} \geq 1$ at $-7.0 < \eta <-6.0$ \textbf{or} $3.7 < \eta < 4.8$ (upper squares)(d).

\emph{}

\begin{figure}[ht]
\centering
   \includegraphics[width=0.48\textwidth]{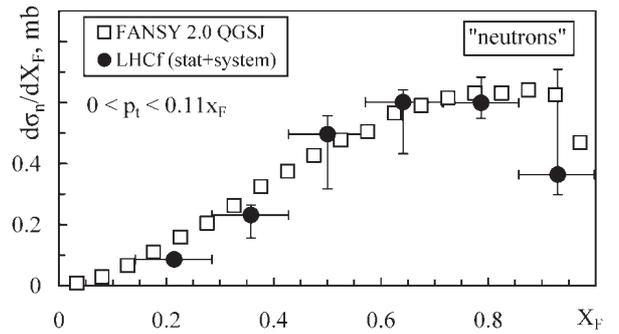}
  \caption{FANSY 2.0 and LHCf "neutron" $d\sigma/dx_F$ spectrum at $p_t < 0.11 x_F$.}
  \label{pp_n_dsigmadXLHCf}
\end{figure}

\begin{figure}[ht]
\centering
   \includegraphics[width=0.48\textwidth]{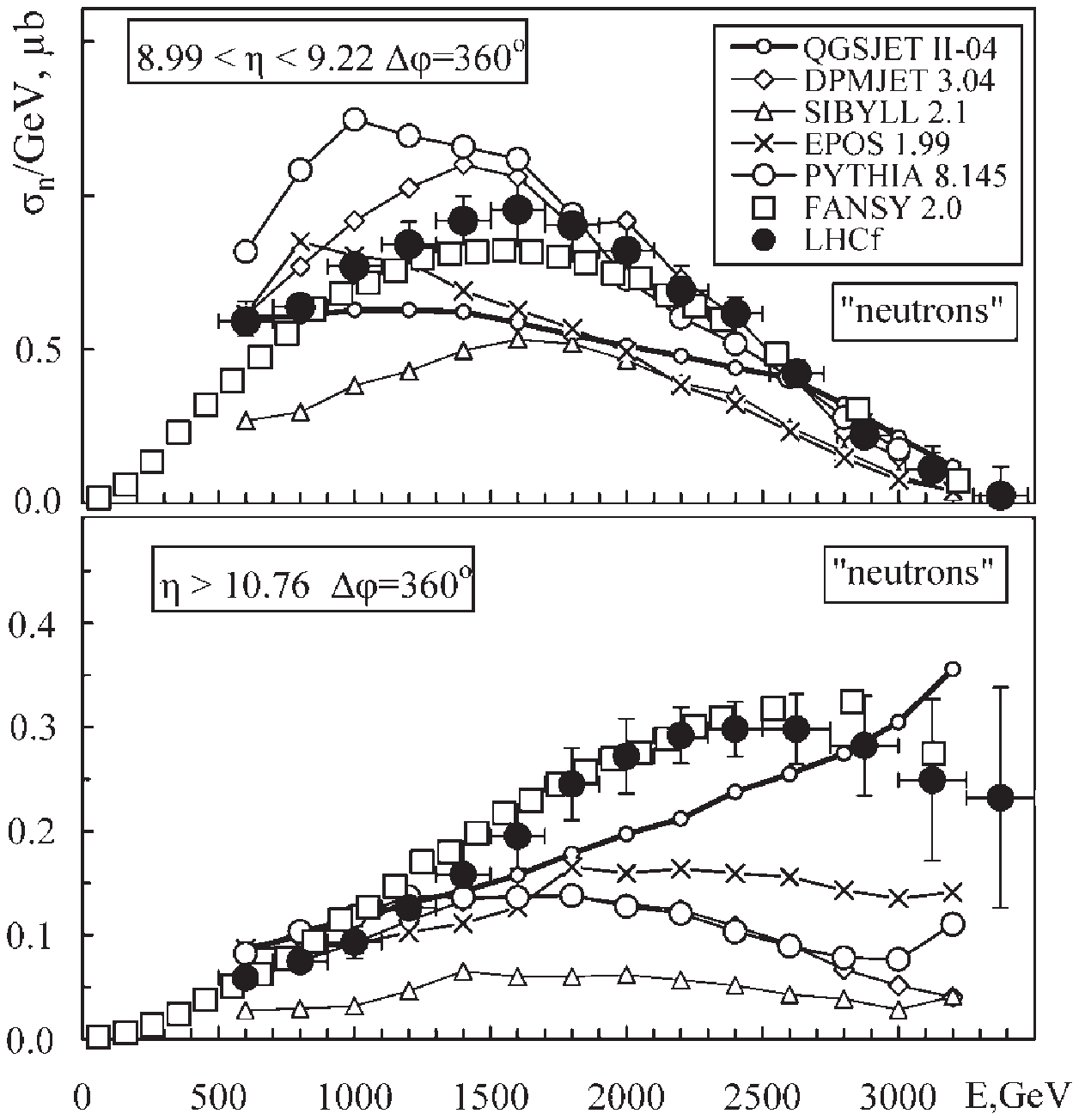}
  \caption{FANSY 2.0 and LHCf "neutron" $d\sigma/dE$ energy spectrum at $8.99 < \eta < 9.2$ (top) and at $\eta > 10.76$ (bottom), $\Delta \phi = 360^\circ $ and $\sqrt{s} = 7$ TeV.}
  \label{pp_neutron_dsdE}
\end{figure}

Fig. \ref{pp_n_dsigmadXLHCf} shows
 FANSY 2.0 and LHCf "neutron" $(n, \bar{n}$, K$_s,L^0$) $d\sigma/dx_F$ spectrum at $p_t < 0.11 x_F$ )  at $\Delta \phi = 360^\circ$ and $\sqrt{s} = 7$ TeV.
Fig. \ref{pp_neutron_dsdE} shows
 LHCf and simulated "neutron" $d\sigma/dE$ energy spectra at $8.99 < \eta < 9.2$ (top) and at $\eta > 10.76$ (bottom).


\begin{figure}[ht]
\centering
   \includegraphics[width=0.48\textwidth]{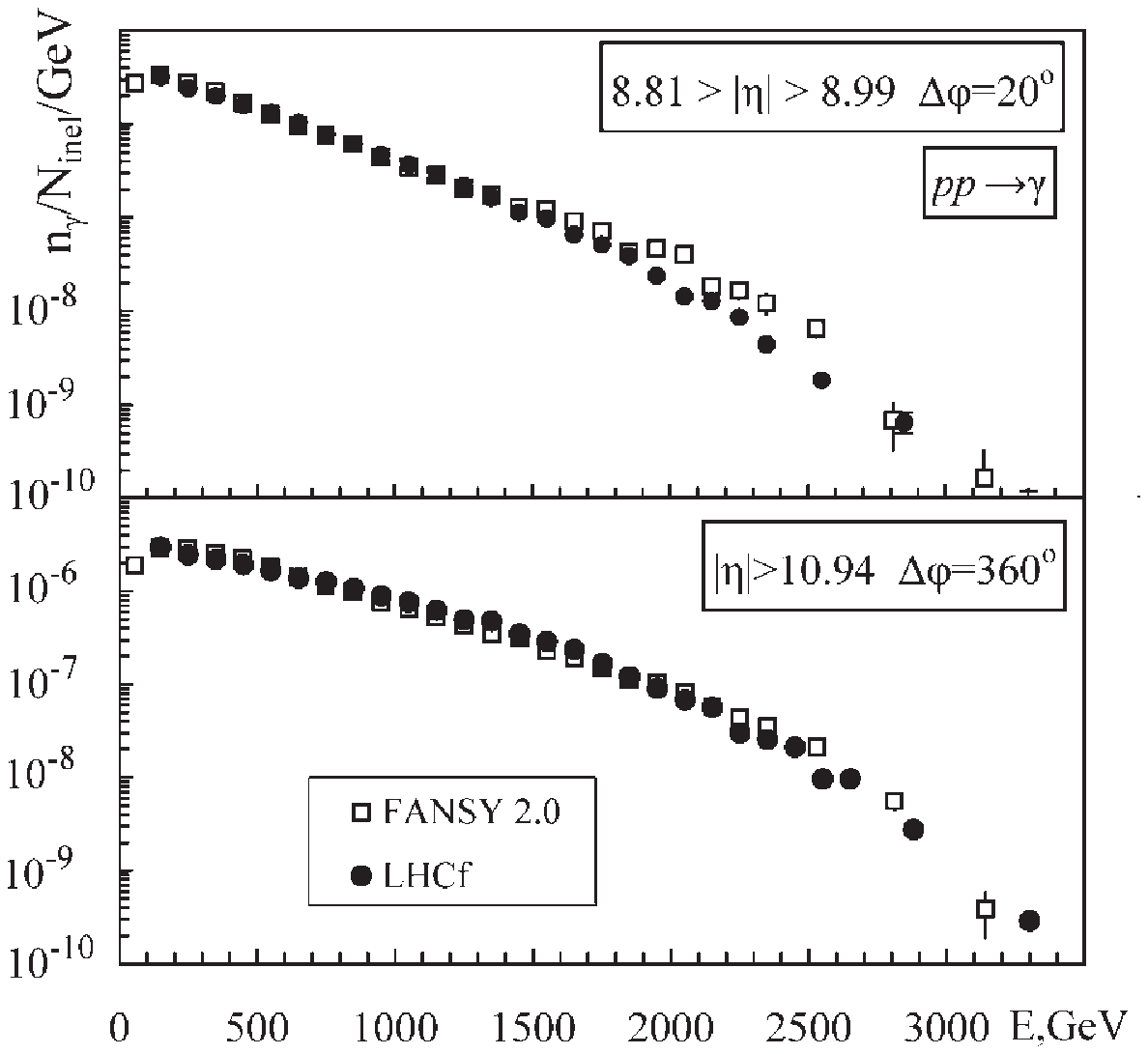}
  \caption{LHCf data and FANSY 2.0 results on $\gamma$-ray $n_\gamma/N_{inel}/$GeV energy spectra at $8.81 < \eta < 8.89$, $\Delta \phi = 20^\circ $  (top) and
  $ \eta > 10.94$, $\Delta \phi = 360^\circ $ (bottom).}
  \label{pp_gamma_dndELHCf}
\end{figure}

Fig. \ref{pp_gamma_dndELHCf} shows FANSY 2.0 and LHCf $\gamma$-ray $n_\gamma/N_{inel}$ energy spectra in the following pseudorapidity and azimuthal-angle ranges  \cite{LHCfPLB2011},
  $8.81 < \eta < 8.89$, $\Delta \phi = 20^\circ $  (top) and
  $ \eta > 10.94$, $\Delta \phi = 360^\circ $ (bottom).

\section{Coplanar particle generation}

\subsection{General view}

A tendency to a coplanarity of most energetic cores of $\gamma$-ray--hadron families observed has been first found by the {\em Pamir} Collaboration   \cite{Pamir4_1,Ivan,Kopenetal,Pamir4_2,Borisetal3} and confirmed later in
 \cite{Xue_etal,strana,strana2,capdev1}.
 The probability $W^{fluct}_{tot}$ for the total set of these experimental results to be produced
by cascade fluctuations is much lower than $10^{-10}$ \cite{MukhJHEP,mukhEPJC2009}.
This result illustrates that strong "forward-physics" interactions at superhigh energies are not well-described with the
quark-gluon string model (QGSM) concept.

The phenomenon is related to hadron-nucleus interactions at
$E_0 \gtrsim 10^{16}$ eV ($\sqrt{s}\gtrsim 4$ TeV) \cite{Boris99NPBPS}, characterized by a large cross section (comparable with $\sigma_{inel}^{pp}$) and
was initially interpreted as a manifestation of
large transverse momenta of most energetic fragmentation-range particles ($X_{Lab} = E/E_0 \gtrsim 0.1$) \cite{MukhJHEP}

This phenomenon could be reproduced in the framework of two concepts as a result of
 a) conservation of the angular momentum of a relativistic fast-rotating quark-gluon string (QGS) stretched between colliding hadrons \cite{wibig};
 b) semihard double diffraction  (SHDID) dissociation and appearance of coplanarity as a result of QGS tension inside the diffraction cluster between a semihardly scattered constituent quark and other spectator quarks of the
projectile hadron and its following rupture \cite{Royzen}  with a lower multiplicity and higher average energy of particles;
 c) the most extraordinary explanation assumes that this phenomenon could be described within the recently proposed hypothesis of "crystal world",  with latticed and anisotropic spatial dimensions and decrease of dimension number with increasing energy  \cite{Anchordoquietal}.

 In this work, the first approach is only considered.

To study  this problem,  FANSY 2.0 QGSCPG version is designed \cite{mukhFANSY20}, which simulates both traditional and  coplanar particle generation. More details of comparison of LHC data with results of simulation with the above-described basic FANSY 2.0 QGSJ version are given in \cite{mukhFANSY20}.

\subsection{Coplanarity concepts}
\label{problems}

\begin{figure}[ht]
\centering
   \includegraphics[width=0.42\textwidth]{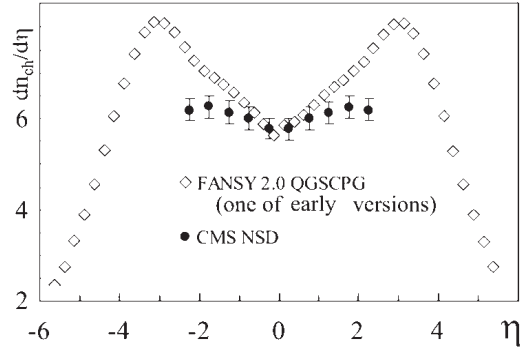}
  \caption{Peaks in $d\sigma / d\eta$ distribution originated by growth of transverse momenta of most energetic particles in primary-concept CPG versions.}
  \label{detamaxcopl}
\end{figure}

\begin{figure}[ht]
\centering
   \includegraphics[width=0.48\textwidth]{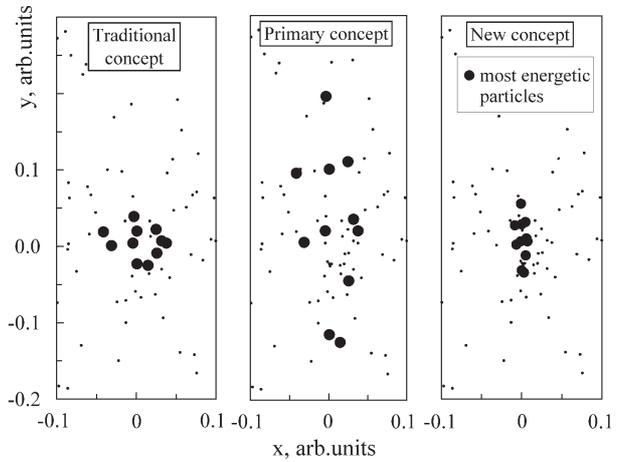}
  \caption{Tracks of particles on the target plane in cases of traditional interaction (left),  coplanar interaction within the primary concept  of increased $p_t$ (middle), coplanar interaction within the new concept of traditional $p_t$ (right). Most energetic particles are shown with large filled circles.}
  \label{targetplan}
\end{figure}

Simulation using tentative FANSY 2.0 QGSCPG versions demonstrated the following fundamental problem.

The originally exploited concept qualitatively explains the observed coplanarity of momenta of most energetic particles with assuming their \textit{high transverse momenta} in a coplanarity plane. However, in this case a significant $p_t$ growth suppresses  $d\sigma / dy$ and $d\sigma / d\eta$ distributions of hadrons at highest $|y|$ and $|\eta |$ values and creates robust peaks at $2 \lesssim |\eta | \lesssim  4$ which are contrary to LHC data (Fig. \ref{detamaxcopl}).
 This was an unsolvable problem for all the QGSCPG versions based on the primary coplanarity concept.
 However, the coplanarity is observed in cosmic-ray experiments. Is it possible to reconcile this result with the LHC data?

Simulation has shown that a general
agreement of LHC data and idea of coplanar generation becomes real only using  a new concept of coplanarity origin, namely, some \textit{decrease of particle's transverse momenta directed normally to the coplanarity plane} takes place so that the
absolute $p_t$ values \textit{do no change}.

Fig. \ref{targetplan} shows three examples of tracks of particles, generated in the same imaginary interaction, on a target plane, placed at some distance from the interaction point, in the cases of traditional QGSM-like interaction (left), primary-concept coplanar interaction with increased $p_t$ (middle), and new-concept coplanar interaction  with traditional $p_t$  (right).  Most energetic  particles are shown with large black circles.
The geometric scale is given in arbitrary units.

\subsection{Coplanarity simulation}
\label{versions}

All simulated interaction characteristics (excluding azimuthal ones)
 simulated with FANSY 2.0  QGSJ and QGSCPG versions, are similar.
 The QGSJ and QGSCPG versions merge smoothly at $\sqrt{s} \lesssim 2$ TeV.

In the QGSCPG version all characteristics of particles are primarily simulated with the traditional way. If the summary energy of secondary particles is higher than a fixed value, transversal momenta of high-rapidity particles  in the coplanarity range ($|y| > y_{copl}$), the algorithm turns $\overrightarrow{p}_t$ of each such particles towards the coplanarity plane. This plane is determined by momenta of the interacting hadrons and $\overrightarrow{p}_t$  of leaders surviving after the collision. The trend of turning of transverse momenta of secondary particles to this plane is weakening at $y_{int} < |y| < y_{copl}$ (in the intermediate range) and disappears at $|y| < y_{int}$ ( the traditional range). Here $|y_{max}| = \sqrt{s}/2 /m$,  $y_{copl} \approx y_{max} - \Delta y \approx 5 -6$, $y_{int}\approx 2 -3$, $\Delta y = 3- 5$, $m$ is particle mass.

\begin{figure}[ht]
\centering
    \includegraphics[width=0.45\textwidth]{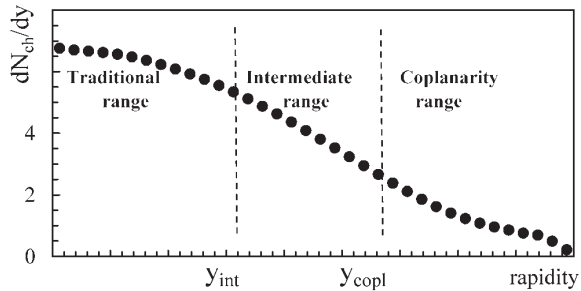}
   \caption{Qualitative dependence of the traditional, intermediate and coplanarity ranges on rapidity.}
   \label{CPGranges}
 \end{figure}

 Fig. \ref{CPGranges} shows a qualitative dependence of the traditional, intermediate and coplanarity ranges on rapidity.

\subsection{On search for coplanarity at LHC}
\label{CASTOR}

\begin{figure}[ht]
\centering
   \includegraphics[width=0.40\textwidth]{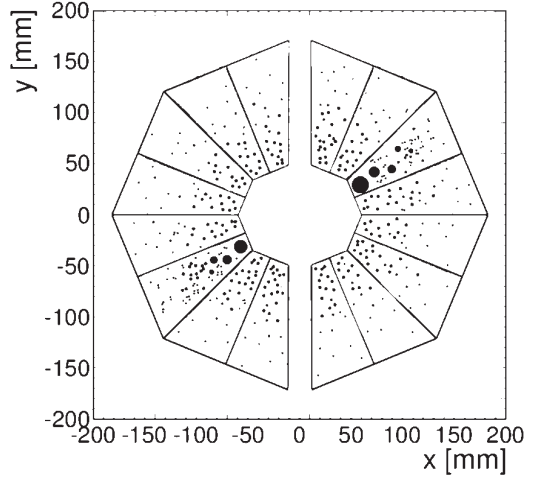}
  \caption{CASTOR's simplified cross section scheme. Black circles show tracks of particles. High-energy particles shown with large circles demonstrate a tendency to coplanarity.}
  \label{CASTORregisterE}
\end{figure}

The CASTOR experiment seems to be promising in study of coplanarity (at least, in the
framework of FANSY 2.0).
 Fig. \ref{CASTORregisterE} shows a simplified CASTOR's cross section scheme and an example of detection of one coplanar interaction.
 The detector consists of 16 segments and is divided in the middle by a vertical slit.
Particles are considered to be detected if their pseudorapidity values are in the range of $5.3 < \eta < 6.5$ and they do
not fall into the vertical slit. Black circles in Fig. \ref{CASTORregisterE} show tracks of particles. The larger the circle size, the higher the energy of the particle. High-energy particles show a tendency to some coplanarity.  Low-energy particles form a more or less azimuthally symmetric halo.

To analyze events, energy values "measured" in each of the segments, $E_i$ ($E_i \geq 100$ GeV, $i$ is the number of a segment), are used. Events
with total energy $\sum E_i \geq 1$ TeV "measured" in two or more segments are only analyzed.
 The first number is assigned to a segment with a maximum release of
energy, $E_{max}$. Here and below, $E_1=E_{max}$, $E_{cop}=E_1 + E_9$;  $E_{tr}=E_5+E_{13}$, i.e., it is the energy measured in  9th and 13th
segments, perpendicular to the first segment. A simple parameter is applied, namely,
 $\varepsilon_{cop} = E_{cop}/(E_{cop}+E_{tr})$, which characterizes the event coplanarity degree.
 If $\varepsilon_{cop} =1$,  the degree of event coplanarity is maximum.

\begin{figure}[ht]
\centering
   \includegraphics[width=0.48\textwidth]{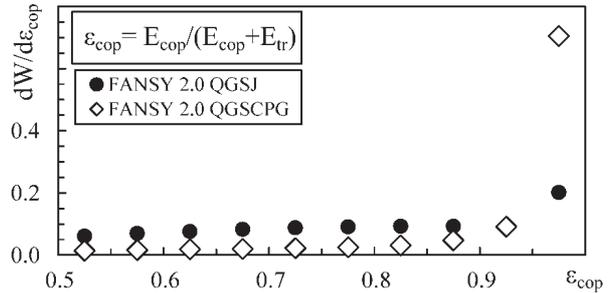}
  \caption{CASTOR's $\varepsilon_{cop}$ distributions for FANSY 2.0 QGSJ and QGSCPG  at $\sqrt{s}=7$ TeV.}
  \label{CASTORepsilcop}
\end{figure}

Fig. \ref{CASTORepsilcop} shows $\varepsilon_{cop}$ distributions for FANSY 2.0 QGSJ and QGSCPG versions at $\sqrt{s}=7$ TeV.  Difference between model predictions becomes very large  at $\varepsilon_{cop} \rightarrow 1$.

\section*{Conclusion}
\label{conclus}
The FANSY 2.0 Monte Carlo code is designed to study superhigh-energy cosmic-ray "forward physics" interactions and includes traditional \textit{QGSJ} QGSM-based version as well as a \textit{QGSCPG} one which realizes a coplanar particle generation (CPG).

CPG process simulated with FANSY 2.0 QGSCPG does not contradict to LHC data and could be tested in the CASTOR experiments.
To reconcile experimental and simulated data, it is necessary to replace the primary concept of \textit{growth} of transverse momenta of most energetic particles in the coplanarity plane with a new concept of \textit{reduction} of transverse momenta directed normally to the coplanarity plane.

A version of appearance of CPG processes in double-diffraction interactions only requires a separate consideration, as in this case energy spectrum of secondary particles is more hard that is important in cosmic-ray experiments.

\end{document}